\documentclass[11pt,showpacs]{revtex4}
\input epsf.sty

\usepackage{ulem}
\usepackage{cancel}
\usepackage{color}

\def\comment#1{}

\def\E{{\mathcal E}}

\usepackage{graphicx}
\usepackage{slashbox}
\begin{document}
\title{Ultraviolet fixed point and massive composite particles in TeV scales}
\author{She-Sheng Xue}
\email{xue@icra.it}
\affiliation{ICRANeT, Piazzale della Repubblica, 10-65122, Pescara,\\
Physics Department, University of Rome ``La Sapienza'', Rome,
Italy} 


\begin{abstract}
We present a further study of the dynamics of high-dimension fermion 
operators attributed to the theoretical inconsistency of the 
fundamental cutoff (quantum gravity) and the parity-violating gauge 
symmetry of the standard model. Studying the 
phase transition from a symmetry-breaking phase to 
a strong-coupling symmetric phase and the $\beta$-function behavior 
in terms of four-fermion coupling strength, we discuss the critical 
transition point as a ultraviolet-stable fixed point where  
a quantum field theory preserving the standard 
model gauge symmetry with composite particles can be realized. 
The form-factors and masses of composite particles at 
TeV scales are estimated by extrapolating the 
solution of renormalization-group equations from the infrared-stable 
fixed point where the quantum field theory of standard model is 
realized and its phenomenology including Higgs mass has been experimentally 
determined. We discuss the probability of composite-particle 
formation and decay that could be experimentally verified 
in the LHC by measuring the invariant mass of relevant final states and their peculiar kinetic distributions.           
\end{abstract}


\maketitle

\noindent
{\bf Introduction.}
\hskip0.1cm
The parity-violating (chiral) gauge symmetries and spontaneous/explicit 
breaking of these symmetries for the hierarchy of fermion masses have been at 
the center of a conceptual elaboration that has played a major role in 
donating to mankind the beauty of the standard model (SM) for particle physics. 
The Nambu-Jona-Lasinio model (NJL) \cite{njl} of four-fermion interactions
at high energies and its effective counterpart, the Higgs model \cite{higgs} of 
fermion-boson Yukawa interactions at low energies, provide an elegant 
description for the electroweak symmetry breaking and intermediate gauge boson 
masses. After a great experimental effort for many years, the 
ATLAS \cite{ATLAS} and CMS \cite{CMS} experiments have recently shown the first observations of a 126 GeV scalar particle in the search for the Standard
Model Higgs boson at the LHC. This far-reaching result
begins to shed light on this most elusive and fascinating arena of fundamental particle physics. 

It is an important issue to study the dynamics at high-energy scale that 
originates the high-dimensional operators of fermion fields.
The strong technicolor dynamics of extended gauge theories at the 
TeV scale was invoked \cite{hill1994,bhl1990a}
to have a natural scheme incorporating the 
relevant four-fermion operator (\ref{bhl}) of the NJL type. 
We here present a 
brief introduction that the origin of high-dimensional operators of all fermion 
fields is due to the quantum gravity at the Planck length 
($a_{\rm pl}\sim 10^{-33}\,$cm, 
$\Lambda_{\rm pl}=\pi/a_{\rm pl}\sim 10^{19}\,$GeV).
Studying the quantum Einstein-Cartan theory in the framework of Regge calculus, we calculated \cite{xue2010} the minimal length 
$(\approx 1.2\,a_{\rm pl})$ of discrete space-time, which provides a natural regulator for local quantum field 
theories of particles and gauge interactions. On the other hand, based on 
low-energy observations of parity violation, the SM Lagrangian was built in 
such a way as to preserve the exact chiral-gauge-symmetries 
$SU_L(2)\otimes U_Y(1)$ that are accommodated by elementary 
left-handed fermions 
and right-handed fermions. However, a profound result, in the form of a 
generic no-go theorem \cite{nn1981}, tells us that there is no 
consistent way to straightforwardly transpose on a discrete space-time 
the bilinear fermion Lagrangian of the continuum SM theory in such a way 
as to exactly preserve the chiral gauge symmetries. We are led to 
consider at least quadrilinear fermion interactions to preserve the 
chiral gauge symmetries. As an example, the four-fermion operator in the Einstein-Cartan theory can be 
obtained by integrating over static torsion fields at the Planck scale 
\cite{torsion}. The very-small-scale structure of space-time and 
high-dimensional operators of fermion fields must be very complex 
as functions of the space-time spacing $\tilde a$ and the gravitational 
gauge-coupling $g_{\rm grav}$ between fermion fields and quantum 
gravity at the Planck scale. 
As the running gravitational 
gauge-coupling $g_{\rm grav}(\tilde a)$ is approaching to its 
ultraviolet (UV) stable critical point $g^{\rm crit}_{\rm grav}$ 
for $\tilde a \rightarrow a_{\rm pl}$ \cite{xue2012}, the physical 
scale $\Lambda = \Lambda[g_{\rm grav}(\tilde a),\tilde a]$ 
($\Lambda^{-1} \gg \tilde a$) satisfies the 
renormalization-group (RG) equation in the scaling region of the 
UV-stable fixed point, where the irrelevant high-dimensional 
operators of fermion fields are suppressed at 
least by ${\mathcal O}(\Lambda/\Lambda_{\rm pl})$; only the relevant 
operators receive anomalous dimensions and become effectively 
renormalizable dimension-4 operators at the high-energy scale $\Lambda$.

On the other hand, these relevant operators can be constructed on the basis of the phenomenology of SM at low-energies.  
In 1989, several authors \cite{nambu1989,Marciano1989,bhl1990} 
suggested that the symmetry breakdown of SM could be 
a dynamical mechanism of the NJL type that 
intimately involves the top quark at the high-energy scale $\Lambda$. 
Since then, many models based on this idea have been 
studied \cite{DSB_review}. The top-quark and Higgs-boson masses 
were supposed to be
achieved by the RG equations in the scaling region 
of the infrared (IR) stable fixed 
point \cite{bhl1990a,Marciano1989,bhl1990}. In the following discussions, we 
adopt the BHL model of 
an effective four-fermion operator \cite{bhl1990}
\begin{eqnarray}
L = L_{\rm kinetic} + G(\bar\psi^{ia}_Lt_{Ra})(\bar t^b_{R}\psi_{Lib}),\quad G\sim 1/\Lambda^2
\label{bhl}
\end{eqnarray}
in the context of a well-defined quantum field theory at the 
high-energy scale $\Lambda$. 

For the reason that the four-fermion
interaction may be due to quantum gravity at the Planck 
scale where all fermions should be on an equal footing, we 
generalized \cite{xue2013_1} the Lagrangian (\ref{bhlx}) to 
\begin{eqnarray}
L &=& L_{\rm kinetic} + G(\bar\psi^{ia}_L\psi_{Rja})(\bar \psi^{jb}_R\psi_{Lib})+{\rm terms},\nonumber\\
&=& L_{\rm kinetic} + G(\bar\psi^{ia}_Lt_{Ra})(\bar t^b_{R}\psi_{Lib})
+ G(\bar\psi^{ia}_Lb_{Ra})(\bar b^b_{R}\psi_{Lib})+{\rm terms},
\label{bhlx}
\end{eqnarray}
where $a,b$ and $i,j$ are the color and flavor indexes 
of the top and bottom quarks, the $SU_L(2)$ doublet 
$\psi^{ia}_L=(t^{a}_L,b^{a}_L)$ 
and the singlet $\psi^{a}_R=t^{a}_R,b^{a}_R$ are the eigenstates 
of the electroweak interaction, and addition terms for 
the first and second quark families can be obtained 
by substituting $t\rightarrow u,c$ and $b\rightarrow d,s$ \cite{xuelepton}. Moreover,
we showed that the less numbers of Goldstone modes (positive energy) 
are, and the smaller total energy of the system is, as a result the minimal dynamical symmetry breaking (\ref{bhl}) 
is an energetically favorable configuration 
(ground state) of the quantum field theory with high-dimension operators of all fermion fields at the 
cutoff $\Lambda$.

It was shown \cite{xue1997,xueprd2000,xuejpg2003} that
if the four-fermion coupling $G(\mu)$ 
is larger than a critical value $G_{\rm crit}$, and the energy scale $\mu$ is larger than a threshold energy scale $\E_{\rm thre}$, the weak-coupling 
symmetry-breaking phase transits to the strong-coupling symmetric 
phase where massive composite particles are formed fully preserving 
the chiral gauge symmetries of SM, and the parity-symmetry is 
restored.
In Ref.~\cite{xue2013}, we found a unique solution to the RG equation 
in the symmetry-breaking phase, which indicates 
the threshold energy scale $\E_{\rm thre}\approx 4.27\,$TeV 
and the form-factor 
of composite Higgs boson $\tilde Z_H(\E_{\rm thre}) \approx 1.1$, 
corresponding to the Higgs-boson mass $m_{_H}\approx 126.7\,$GeV and 
top-quark mass $m_t\approx 172.7\,$GeV. As a consequence, these masses and  
the pseudoscalar decay constant $f_\pi$ can be obtained without drastically 
fine-tuning the four-fermion coupling. 

In this Letter, utilizing the BHL model (\ref{bhl}) 
in the symmetry-breaking phase, we numerically solve the 
RG equations of the SM with an infrared boundary conditions 
fixed by the top-quark and Higgs-boson masses recently 
measured, and obtain the form-factor 
of composite Higgs boson, increasing as the energy scale $\mu$ 
increasing up to the energy threshold $\E \approx 5\,$TeV, at which the 
Higgs-boson 
quartic coupling $\bar \lambda(\E)$ vanishes. 
This is different from the BHL result 
obtained by imposing the compositeness conditions of 
the form-factor vanishing at high-energy 
cutoff scale $\Lambda$.   
Moreover, we show that in the symmetry-breaking phase the $\beta(G)$-function
is positive near to an infrared-stable fixed point for the SM, while 
the $\beta(G)$-function is negative in the strong-coupling symmetric phase, 
where the composite Higgs boson combines 
with an elementary  fermion to form a massive composite fermion. 
This implies that the critical point of the second-order phase transition 
should be a UV-stable fixed point. The result $\E\approx 5\,$TeV from the 
solution to RG-equations
infers the energy scale in the scaling region of the UV-stable fixed point.  
As a result, we estimate the 
spectra of massive composite particles and discuss the 
high-energy collider signatures of these composite particles, 
which could be identified by the resonance in invariant mass and particular
kinematic distribution of final states measured.  

\begin{figure}[t]
\begin{center}
\includegraphics[height=1.25in]{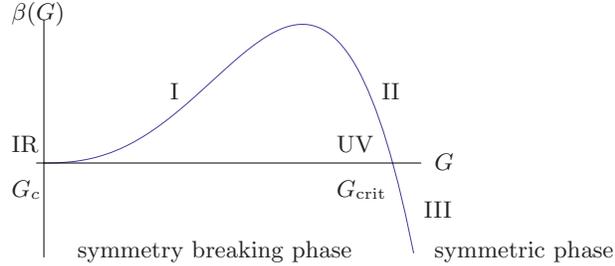}
\put(-155,90){\footnotesize $\beta(G)$}
\put(5,33){\footnotesize $G$}
\put(-130,00){\footnotesize symmetry breaking phase}
\put(5,00){\footnotesize symmetric phase}
\put(-95,60){\footnotesize I}
\put(-15,60){\footnotesize II}
\put(1,15){\footnotesize III}
\put(-155,23){\footnotesize $G_c$}
\put(-155,40){\footnotesize ${\rm IR}$}
\put(-32,23){\footnotesize $G_{\rm crit}$}
\put(-32,40){\footnotesize ${\rm UV}$}
\caption{This is a sketch to qualitatively show the behavior of the 
$\beta$-function in terms of the four-fermion coupling $G$. We indicate 
the quasi IR-stable fixed point $G_c$ and a possible UV-stable fixed 
point $G_{\rm crit}$, the latter separates the symmetry-breaking phase 
(positive $\beta(G)$-function) from the symmetric phase (negative 
$\beta(G)$-function). We also indicate the 
positive parts  ``I'' (increasing) and ``II'' (decreasing), as well as 
the negative part ``III'' of the $\beta(G)$-function.} \label{fixp}
\end{center}
\end{figure}

\noindent
{\bf The IR-stable fixed point and symmetry-breaking phase.}
\hskip0.1cm 
In this phase, the quantum field theory (\ref{bhl}) contains the 
massive spectra of top quark and composite Higgs boson.
Employ the ``large $N_c$-expansion'', i.e., keep $GN_c$ fixed and 
construct the theory 
systematically in powers of $1/N_c$. At the lowest order of one
fermion-loop contribution, one obtains the gap equation 
for top-quark mass $m_t\not=0$ 
\begin{eqnarray}
\frac{1}{G_c}-\frac{1}{G}=\frac{1}{G_c}\left(\frac{m_t}{\E}\right)^2
\ln \left(\frac{\E}{m_t}\right)^2>0,
\label{delta}
\end{eqnarray}
for $G\gtrsim G_c\equiv 8\pi^2/(N_c\E^2)$, where 
$\E\approx \E_{\rm thre}$ characterizes the energy scale of restoring 
symmetries for $G\gtrsim G_{\rm crit}$. 
In Eq.~(\ref{delta}), considering $m_t$ as a running energy 
scale $\mu$, we can approximately obtain the running coupling
\begin{eqnarray}
G(\mu)\approx G_c\left[1-\left(\frac{\mu}{\E}\right)^2
\ln \left(\frac{\E}{\mu}\right)^2\right]^{-1},
\label{runn}
\end{eqnarray}
and the $\beta$-function 
\begin{eqnarray}
\beta(G)\equiv \mu \frac{dG}{d\mu} 
&\approx & 2\frac{G^2}{G_c}\left(\frac{\mu}{\E}\right)^2
\left[1+\ln \left(\frac{\E}{\mu}\right)^2\right]>0,
\label{gbeta}
\end{eqnarray}
for $G_{\rm crit}> G \gtrsim G_c $ and 
$\E > \mu \gtrsim v$, where $v$ is the electroweak scale. 
The positive $\beta$-function of Eq.~(\ref{gbeta}) 
indicates that $G_c$ is an IR-stable fixed point, 
$G\rightarrow G_c+0^+$ as $\mu\rightarrow v$. Here
we ignore the behavior of the functions $G(\mu)$ and $\beta(G)$ for 
$G\rightarrow G_c+0^-$ in the weak-coupling symmetric phase ($G<G_c$), 
therefore $G_c$ should be regarded as a ``quasi'' IR-stable fixed point. 
To represent the behavior of the $\beta(G)$-function
discussed up to now, we sketch in Fig.~\ref{fixp} the positively 
increasing curve ``I'' of the $\beta(G)$-function departing 
from $G=G_c$, where the coupling $G(\mu)$ increases as the energy 
scale $\mu$ increases in the range $v\lesssim \mu <\E$.

\begin{figure}
\begin{center}
\includegraphics[height=1.25in]{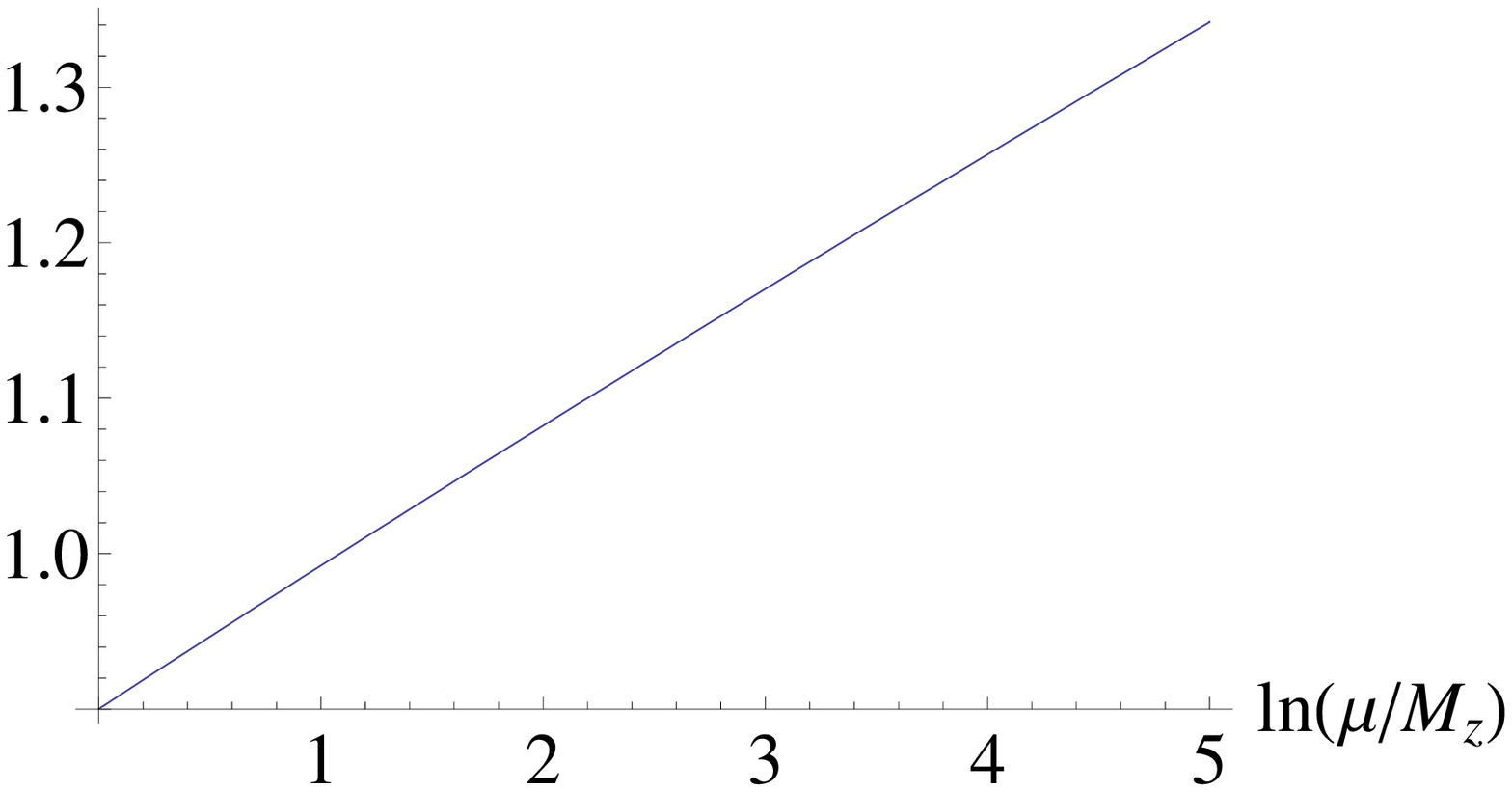}
\includegraphics[height=1.25in]{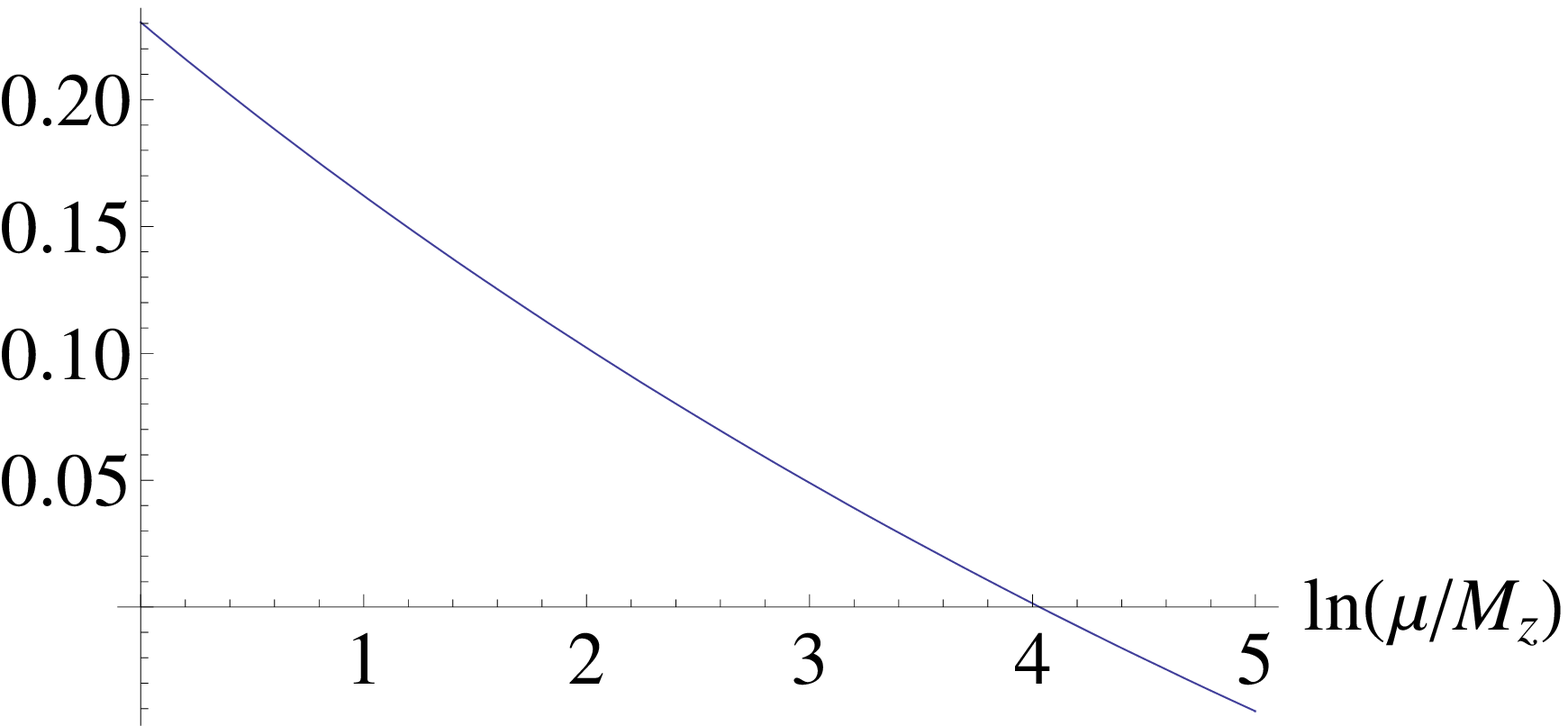}
\put(-370,95){\footnotesize $\tilde Z_{H}(\mu)$}
\put(-180,95){\footnotesize $\tilde \lambda(\mu)$}
\caption{Using all experimentally measured quantities at low energies, we numerically solve the RG 
equations (\ref{reg1}), (\ref{reg2}) and boundary conditions 
(\ref{thmass}), (\ref{varmass}) to uniquely determine the functions 
$\tilde Z_{H}(\mu)$ and $\tilde\lambda(\mu)$ of Eq.~(\ref{boun0}) in terms of 
the energy scale $\mu> M_z$. Since $\tilde\lambda(\E)$ cannot be negative, 
otherwise the total energy of the system would not be bound from below, 
we numerically determine the values (\ref{thvari}) of 
$\E$ and $\tilde Z_{H}(\E)$ by demanding $\tilde\lambda(\E)=0$.} \label{figrg}
\end{center}
\end{figure}

\noindent
{\bf The scaling region of the IR-stable fixed point.}
\hskip0.1cm 
The full induced effective Lagrangian of the low-energy SM 
in the scaling region of the 
IR-stable fixed point takes the form \cite{bhl1990}
\begin{eqnarray}
L &=& L_{\rm kinetic} + g_{t0}(\bar \Psi_L t_RH+ {\rm h.c.})
+\Delta L_{\rm gauge}\nonumber\\ 
&+& Z_H|D_\mu H|^2-m_{_H}^2H^\dagger H
-\frac{\lambda_0}{2}(H^\dagger H)^2,
\label{eff}
\end{eqnarray} 
and all renormalized quantities received fermion-loop contributions are 
defined with respect to the low-energy scale $\mu$. 
The conventional renormalization $Z_\psi=1$ for fundamental 
fermions and the unconventional wave-renormalization (form factor)
$\tilde Z_H$ for composite Higgs bosons are 
adopted \cite{bhl1990} 
\begin{equation}
\tilde Z_{H}(\mu)=\frac{1}{\bar g^2_t(\mu)},\, \bar g_t(\mu)=\frac{Z_{HY}}{Z_H^{1/2}}g_{t0}; \quad \tilde \lambda(\mu)=\frac{\bar\lambda(\mu)}{\bar g^4_t(\mu)},\,\bar\lambda(\mu)=\frac{Z_{4H}}{Z_H^2}\lambda_0,
\label{boun0}
\end{equation}
where $Z_{HY}$ and $Z_{4H}$ are proper renormalization constants of 
the Yukawa-coupling and quartic coupling in Eq.~(\ref{eff}). 
In the scaling region of the IR-stable fixed point 
where the SM 
of particle physics is realized, we utilize the full one-loop RG equations for running couplings $\bar g_t(\mu^2)$ and $\bar \lambda(\mu^2)$
\begin{eqnarray}
16\pi^2\frac{d\bar g_t}{dt} &=&\left(\frac{9}{2}\bar g_t^2-8 \bar g^2_3 - \frac{9}{4}\bar g^2_2 -\frac{17}{12}\bar g^2_1 \right)\bar g_t,
\label{reg1}\\
16\pi^2\frac{d\bar \lambda}{dt} &=&12\left[\bar\lambda^2+(\bar g_t^2-A)\bar\lambda + B -\bar g^4_t \right],\quad t=\ln\mu \label{reg2}
\end{eqnarray}
where one can find $A$, $B$ and RG equations for 
running gauge couplings $g^2_{1,2,3}$ in Eqs.~(4.7), (4.8) of 
Ref.~\cite{bhl1990}. 
In this IR scaling region, the 
electroweak scale $v\approx 239.5\,$GeV and the mass-shell conditions
\begin{eqnarray}
m_t=\bar g_t(m_t)v/\sqrt{2},\quad m_{_H}^2/2=\tilde \lambda (m_{_H}) v^2,
\label{thmass}
\end{eqnarray}
are set in.
Using the experimental values
of $M_w$, $M_z$, $g^2_{1,2,3}$, $\cdot\cdot\cdot$ including 
the top-quark and Higgs-boson masses, 
\begin{eqnarray}
m_{_H}=126\pm 0.5 \,{\rm GeV};\quad m_t=172.9\pm 0.8\, {\rm GeV},
\label{varmass}
\end{eqnarray}
we adopt (\ref{thmass}) as an infrared boundary condition  
to integrate the RG equations 
(\ref{reg1}) and (\ref{reg2}) so as to uniquely determine 
the functions of $\tilde Z_{H}(\mu)$ 
and $\tilde\lambda(\mu)$ (see Fig.~\ref{figrg}), as well as the values of $\tilde Z_{H}(\E)$ 
and the energy scale $\E$ for $\tilde\lambda(\E)=0$. 
We examine the variations of $\tilde Z_{H}(\E)$ 
and $\E$ values corresponding to the uncertainties in experimental 
measurements (\ref{varmass}). 
The results are reported in Tab.~\ref{exp_error} and 
the maximal variations are 
\begin{eqnarray}
\E =5.1 \pm 0.7\,\, {\rm TeV},\quad \tilde Z_H =1.26\pm 0.02.
\label{thvari}
\end{eqnarray}
This indicates that as a unique solution to the 
RG equations (\ref{reg1}) and (\ref{reg2}),
how much variations of $\E$ and $\tilde Z_{H}(\E)$ in high energies  
correspond to the variations of boundary values (\ref{thmass}) in low energies, due to the uncertainties of top-quark and Higgs-boson masses 
(\ref{varmass}).  Note that the uncertainties of gauge couplings and boson masses have not been taken in account in this calculations.

It is important to compare and contrast our study with the BHL one 
\cite{bhl1990}. In both studies, 
the definitions of all physical quantities are identical, the same RG 
equations (\ref{reg1}) and (\ref{reg2}) are used for running Yukawa 
and quartic couplings as well as gauge couplings. However, 
the different boundary conditions are adopted. 
We impose the infrared boundary condition (\ref{thmass}) with 
(\ref{varmass}) that are known 
nowadays, to uniquely determine 
the solution of the RG equations, and values of the form-factor
$\tilde Z_{H}(\E)\not=0$ and high-energy scale 
$\E\, [\tilde \lambda(\E)=0]$,  
as shown in Fig.~\ref{figrg} $\tilde Z_{H}(\mu)$ [$\tilde \lambda(\mu)$] 
monotonically increases (decreases) as the energy scale $\mu$ increases up to 
$\E$. 
Both experimental $m_t$ and $m_{_H}$ values were 
unknown in the early 1990s, in order to find low-energy values 
$m_t$ and $m_{_H}$ close to the IR-stable fixed point, 
BHL \cite{bhl1990} imposed the 
compositeness conditions $\tilde Z_{H}(\Lambda)=0$ and 
$\tilde \lambda(\Lambda)=0$ for different values of
the high-energy cutoff $\Lambda$ as the boundary condition to 
solve the RG equations. As a result,  
$m_t$ and $m_{_H}$ values (Table I in Ref.~\cite{bhl1990}) were obtained, and 
we have reproduced these values. However,
these BHL results are radically different from the present results of 
Eqs.~(\ref{varmass}), (\ref{thvari}) and 
Fig.~\ref{figrg}, showing that the composite Higgs boson actually 
becomes a more and more
tightly bound state, as the energy scale $\mu$ increases, and eventually 
combines with an elementary fermion to form a composite fermion 
in the symmetric phase (see next section). This phase transition to 
the symmetric phase is also indicated by $\tilde \lambda(\mu)\rightarrow 0^+$ as $\mu\rightarrow \E+0^-$ at which 
the 1PI vertex function $Z_{4H}$ in Eqs.~(\ref{boun0}), (\ref{eff})
vanishes.

The Yukawa coupling $\bar g_t(\mu) = [\tilde Z_H(\mu)]^{-1/2} < 1$ 
and quartic coupling 
$\bar\lambda(\mu)=\tilde\lambda(\mu)\bar g^4_t(\mu)< 0.15$ 
for $m_{_H}< \mu <\E$, 
as shown in Fig.~\ref{figrg}.  
This consistently indicates that the 
RG equations (\ref{reg1}) and (\ref{reg2})  
derived from perturbative calculations for small couplings 
$\bar g_t(\mu)$ and $\bar \lambda(\mu)$ are
reliable to obtain the numerical results 
$\tilde Z_H(\E)$ and $\E$ of Eq.~(\ref{thvari}). The non-vanishing form-factor
$\tilde Z_H(\mu)$ means that after conventional wave-function and vertex 
renormalizations $Z^{1/2}_H H\rightarrow H$, 
$Z_{HY}g_{t0}\rightarrow g_{t0}$ and 
$Z_{4H}\lambda_{0}\rightarrow \lambda_{0}$ 
[see Eqs.~(\ref{eff}) and (\ref{boun0})], 
the composite Higgs boson behaves as an elementary particle. 
However, its effective Yukawa coupling $\bar g_t(\mu)$
and quartic coupling $\bar\lambda(\mu)$ 
decrease with the energy scale $\mu$ increasing 
in the range $m_{_H}< \mu <\E$.    
This would have 
some effects on the rate or cross-sections 
of the composite Higgs boson decay or 
other relevant processes. In future work, it will be examined by comparison to electroweak precision data if 
these effects could be low-energy collider signatures 
that would tell this scenario apart from 
the SM with an elementary Higgs boson. 

\begin{table}[t]
\begin{center}
\begin{tabular}{|c|c|c|c|}
\hline \backslashbox[15mm]{$m_{_H}$}{$m_t$}&$172.9+0.8 {\,\rm GeV}$ & $172.9-0.8{\,\rm GeV}$ & $172.9 {\,\rm GeV}$  \\
\hline  $126+0.5 {\,\rm GeV}$
& $\tilde Z_H=1.24; \E=4.9{\,\rm TeV}$     & $\tilde Z_H=1.28; \E=5.8{\,\rm TeV}$ & $\tilde Z_H=1.26; \E=5.3{\,\rm TeV}$ \\
\hline 
$126-0.5 {\,\rm GeV}$  & $\tilde Z_H=1.24; \E=4.6{\,\rm TeV}$
& $\tilde Z_H=1.28; \E=5.4{\,\rm TeV}$ &  $\tilde Z_H=1.27; \E=5.0{\,\rm TeV}$\\
\hline
$126{\,\rm GeV}$ & $\tilde Z_H=1.24; \E=4.8{\,\rm TeV}$
& $\tilde Z_H=1.28; \E=5.7{\,\rm TeV}$ & $\tilde Z_H=1.26; \E=5.1{\,\rm TeV}$ \\
\hline
\end{tabular}
\end{center}
\caption{The center values of top-quark and Higgs masses are chosen as 
$m_t=172.9 {\,\rm GeV}$ and $m_{_H}=126 {\,\rm GeV}$. This table shows the 
variations of the theoretical values of $\tilde Z_{H}(\E)$ and
$\E$, corresponding to the variations of experimental values of top-quark and 
Higgs masses (\ref{varmass}).} \label{exp_error}
\end{table}
 
\noindent
{\bf The UV-stable fixed point and strong-coupling symmetric phase.}
\hskip0.1cm 
From the results (\ref{thvari}), we can have 
some insight into the energy threshold $\E_{\rm thre}$ and the form-factor 
$\tilde Z_H(\E_{\rm thre})$ of composite particles 
in the strong-coupling symmetric phase, where the composite Higgs boson 
and an elementary fermion are bound to form a three-fermion state to restore 
the symmetry.
In the strong-coupling limit $Ga^{-2}\gg 1$, where 
$a\equiv (\pi/\Lambda)$, the theory (\ref{bhlx}) is in the 
strong-coupling symmetric phase \cite{xue1997,xueprd2000}. This was shown
by scaling
$
\psi(x)\,\rightarrow \psi(x)=a^2 g^{1/4}\psi(x)$ and $g\equiv G/a^4\,
$ ($ga^2\gg 1$),
writing the action (\ref{bhlx}) as
\begin{eqnarray}
S_{\rm kinetic}&=&{1\over 2ag^{1/2}}\sum_{x,\mu}\bar\psi(x)
\gamma_\mu \partial^\mu\psi(x),\quad \partial^\mu\equiv \delta_{x,x+a_\mu}-\delta_{x,x-a_\mu}\label{rfa}\\
S_{\rm int}&=&\sum_{x}\left[(\bar\psi^{ia}_Lt_{Ra})(\bar t^b_{R}\psi_{Lib}) + (\bar\psi^{ia}_Lb_{Ra})(\bar b^b_{R}\psi_{Lib})\right],\label{rs2}
\end{eqnarray}
and using the strong 
coupling (hopping) expansion in powers of $1/g^{1/2}$ to calculate 
two-point functions of composite fermion and boson fields. Using the first 
term ($t_{Ra}$-channel) in Eq.~(\ref{rs2}), 
in the lowest non-trivial order (one-hopping step) 
we obtained (see Section 4 in Ref.~\cite{xue1997}) the propagator of
the composite Dirac fermions: $SU_L(2)$-doublet ${\bf\Psi}^{ib}_D=(\psi^{ib}_L, {\bf\Psi}^{ib}_R)$ and $SU_L(2)$-singlet ${\bf\Psi}^{b}_D=({\bf\Psi}^b_{L},t_R^b)$, where the renormalized composite three-fermion states are:
\begin{equation}
{\bf\Psi}^{ib}_R=(Z^{^S}_R)^{-1}(\bar\psi^{ia}_Lt_{Ra})t^b_{R}\,;\quad {\bf\Psi}^b_{L}=(Z^{^S}_L)^{-1}(\bar\psi^{ia}_Lt_{Ra})\psi^{b}_{iL},
\label{bound}
\end{equation}
with mass $M=2ga$ and form-factor $Z^{^S}_{R,L}=Ma$, the latter is a 
generalized wave-function renormalization of composite fermion operators. 
The composite bosons ( $SU_L(2)$-doublet) are  (see Section 5 in 
Ref.~\cite{xue1997})
\begin{eqnarray}
H^i=[Z^{^S}_H]^{-1/2}(\bar\psi^{ia}_Lt_{Ra}),\quad \mu^2_{_H}= {4\over N_c}\left(g-{2N_c\over a^2}\right),
\label{boundb}
\end{eqnarray} 
where $[Z^{^S}_H]^{1/2}$ and $\mu_{_H}$ respectively are the form-factor 
and mass of composite bosons. Eq.~(\ref{boundb}) confirms 
the spontaneous symmetry breaking $SU(2)\rightarrow U(1)$ by
the effective mass term $\mu^2_{_H}HH^{\dagger}$ changing its sign 
from $\mu^2_{_H}>0$ to $\mu^2_{_H}<0$, $\mu^2_{_H}=0$ gives rise to the critical
coupling $G_{\rm crit}$, whose exact value however has to be 
calculated by non-perturbative numerical simulations.
In the lowest non-trivial order of the strong-coupling expansion,
the positive contribution to the 1PI vertex of the self interacting
term $(HH^\dagger)^2$ is suppressed by $(1/g)^2$. 
Note that the same calculations 
based on the second term ($b_{Ra}$-channel) in Eq.~(\ref{rs2}) 
lead to the composite particles represented by 
Eqs.~(\ref{bound})-(\ref{boundb}) 
with the replacement $t_{Ra}\rightarrow b_{Ra}$, carrying the 
different quantum numbers of the $U_Y(1)$ gauge group. 
These discussions are also the same 
for the first and second quark families 
by substituting the $SU_L(2)$ doublet $(t_{La}, b_{La})$ 
into $(u_{La}, d_{La})$ or $(c_{La}, s_{La})$ and singlet 
$t_{Ra}$ into $u_{Ra}$ or $c_{Ra}$, as well as singlet 
$b_{Ra}$ into $d_{Ra}$ or $s_{Ra}$ in Eq.~(\ref{bhlx}). 

In the symmetry breaking phase and the scaling region 
of the IR-stable fixed 
point, we know the symmetries, particle spectrum 
(fermions and bosons) and all relevant renormalizable operators 
of the SM at low energies [see Eq.~(\ref{eff})]. 
In the strong-coupling symmetric phase,
the three-fermion states (\ref{bound}) are the bound states 
of the composite boson $H^i=(\bar\psi^{ia}_Lt_{Ra})$ and elementary 
fermion $t^b_{R}$ ($\psi^{b}_{iL}$), and the SM 
chiral-gauge symmetries are fully 
preserved by the massive composite fermions ${\bf\Psi}^{ib}_D$ and
${\bf\Psi}^{b}_D$, as well as their vector-like couplings to 
$\gamma,\, W^\pm$, $Z^0$ and gluon 
gauge bosons, consequently leading to the parity-symmetry restoration. 
 
\begin{figure}
\begin{center}
\includegraphics[height=1.25in]{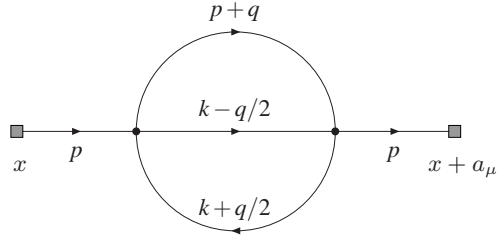}
\put(-13,25){\footnotesize $x+a_\mu$}
\put(-170,25){\footnotesize $x$}
\caption{The Feynman diagram represents the contribution $\sigma(p)$ 
from the one-hopping step of each fermion field in Eq.~(\ref{bound}).} 
\label{hopping}
\end{center}
\end{figure}

We attempt to discuss the possible behaviors ``II'' and ``III'' 
of the $\beta(G)$-function  in the strong-coupling 
regimes (see Fig.~\ref{fixp}). 
To see how the strong coupling $g$ depends on the energy-momentum, 
we need to calculate the corrections from 
more ``hopping'' steps to 
the form-factor ($Z^{^S}_{R,L}=Ma$) and mass ($M=2ga$)
of composite fermions (\ref{bound}). 
In the analogy of calculations presented 
in the Appendix B of 
Ref.~\cite{xue1997} and discussions presented in 
Ref.~\cite{xueprd2000}, 
these corrections
can be approximately calculated by using the train approximation for each fermion of Eq.~(\ref{bound}),
\begin{eqnarray}
&&[Z^{^S}_{R,L}(p)/aM]^{-1} \approx (1+\sigma+\sigma
\sigma+\cdot\cdot\cdot)^3
=\left({1\over 1-\sigma}\right)^3,
\label{train}\\
&&\sigma (p) =-
\frac{2}{(g^{1/2})^4}\left(\frac{\gamma_\nu p^\nu}{p^2}\right)
\!\int^\Lambda_{k,q} 
\frac{\gamma_\mu(p\!+\!q)^\mu}{(p+q)^2} {(k^2-q^2/4)\over(k-q/2)^2(k+q/2)^2},
\label{az1}
\end{eqnarray}
where $p$ is the energy-momentum of composite particles and 
$\sigma(p)$ is represented by Fig.~\ref{hopping}
and its negative sign is attributed to {\it two}
fermion loops. We rewrite Eq.~(\ref{az1}) as
\begin{eqnarray}
\sigma (p)= -\frac{2}{(g^{1/2})^4}\left(\frac{\gamma_\nu p^\nu}{p^2}\right)\gamma_\mu p^\mu \Lambda^4\Phi(p^2/\Lambda^2)=-
\frac{2\pi^4}{G^2\Lambda^4}\Phi(p^2/\Lambda^2),
\label{az2}
\end{eqnarray}
where the dimensionless function $\Phi(p^2/\Lambda^2)$ is a Lorentz scalar. 
Numerical calculations confirm that the function $\Phi(p^2/\Lambda^2)$ is 
positive and finite, monotonically decreases as $p^2/\Lambda^2$ 
increases. As a result, the corrected form-factor 
$Z^{^S}_{R,L}(p)=Ma[1-\sigma(p)]^3$,
leading to the effective running coupling
\begin{eqnarray}
{\mathcal G}(p)\approx  {\mathcal G}\left[1+\frac{6}{{\mathcal G}^2} \Phi(p^2/\Lambda^2)\right], \quad {\mathcal G}\equiv G\times (\Lambda/\pi)^2
\label{ren}
\end{eqnarray}
and the $\beta$-function
\begin{eqnarray}
\beta(G) = p^2\frac{\partial {\mathcal G}(p)}{\partial p^2}\approx \frac{6}{{\mathcal G}} \frac{\partial \Phi(p^2/\Lambda^2)}{\partial \ln(p^2/\Lambda^2)}< 0.
\label{bren}
\end{eqnarray}
This result indicates a negative $\beta$-function and 
$\beta\rightarrow 0^{-}$ in the strong-coupling limit.
Recall  
that in the QED 
case the analogous contribution of {\it one} fermion loop to 
the wave-function 
renormalization constant $Z_3$ is positive, the $\beta$-function 
is positive, i.e., $\beta_{\rm QED}\approx e^3/12\pi^2> 0$. 
On the basis of the $\beta(G)$-function being positive and negative 
respectively in the weak-coupling symmetry-breaking phase 
and strong-coupling symmetric phase, as sketched as ``I'' and ``III'' in 
Fig.~\ref{fixp}, we infer there must be at least one 
zero-point of the $\beta(G)$-function, i.e., $\beta(G_{\rm zero})=0$ and 
$\beta'(G_{\rm zero})<0$. At this zero-point $G_{\rm zero}$,
the positive $\beta(G)$-function ``II'' turns to the negative 
$\beta$-function ``III''. This zero-point $G_{\rm zero}$ is a UV-stable 
fixed point. 

\noindent
{\bf The scaling region of the UV-stable fixed point.}
\hskip0.1cm 
We are not able to determine $G_{\rm zero}$, 
however we expect $G_{\rm zero}\simeq G_{\rm crit}$ for the 
reason that a UV-stable fixed point should be the candidate 
of critical point $G_{\rm crit}$ for 
the second-order phase transition. It is known 
that in the neighborhood of the critical point 
$G_{\rm crit}$, the correlation length $\xi/a$ of the theory goes to infinity, 
leading to the scaling invariance, i.e., the 
renormalization-group invariance. In this scaling region, the running coupling 
$G(a/\xi)$ can be expanded as a series,
\begin{eqnarray}
G(a/\xi)&=& G_{\rm crit} \left[1+ a_0(a/\xi)^{1/\nu}+ {\mathcal O}[(a/\xi)^{2/\nu}]\right]\rightarrow G_{\rm crit}+0^+,
\label{gexp}
\end{eqnarray}
for $a/\xi\ll 1$, leading to the $\beta$-function 
\begin{eqnarray}
\beta(G)&=& (-1/\nu) (G-G_{\rm crit}) + {\mathcal O}[(G-G_{\rm crit})^2]<0\,.
\label{betaexp}
\end{eqnarray}
The correlation length $\xi$ follows the scaling law
\begin{eqnarray}
\xi &= & c_0 a\exp \int^G\frac{dG'}{\beta(G')}
=\frac{ c_0a}{(G-G_{\rm crit})^\nu},
\label{xivary}
\end{eqnarray}
where the coefficient $c_0=(a_0G_{\rm crit})^\nu$ and critical 
exponent $\nu$ need to be determined by 
non-perturbative numerical simulations. Analogously to the electroweak scale 
$v=239.5\,$GeV sets in the scaling region of the IR-stable fixed point $G_c$,
the physical scale $\E_\xi\equiv \xi^{-1}$ sets in the scaling region of 
the UV-stable fixed point $G_{\rm crit}$. This implies the masses of composite particles
\begin{eqnarray}
{\mathcal M} \approx \E_\xi=\xi^{-1},
\label{mcom}
\end{eqnarray} 
and the running coupling  
$G(\mu)|_{\mu\rightarrow\E_{\rm thre}+0^+}\rightarrow G_{\rm crit}$, 
\begin{eqnarray}
G(\mu)\simeq G_{\rm crit}\left[1-\frac{1}{\nu}\ln\left(\frac{\mu}{\E_\xi}\right)\right]^{-1}, \quad \mu/\E_\xi=\xi/(aa_0^\nu) >1\,,
\label{runnc}
\end{eqnarray}
and the scale $\mu$ indicates the energy transfer between constituents 
inside composite particles.
In the scaling region of the UV-stable fixed point, 
all one-particle-irreducible (1PI) functions $\Gamma[\mu, G(\mu)]$ 
of the quantum field theory 
(\ref{bhlx}) at the high-energy scale $\Lambda$ 
evolve to irrelevant or relevant 1PI functions,
as the energy scale $\mu$ increases. The irrelevant 1PI 
functions are suppressed by powers of $(\E_\xi/\Lambda)^n$ and thus 
decouple from the theory. Instead, the relevant 1PI
functions follow the scaling law, therefore are 
effectively dimension-4 and renormalizable, for example 
the propagators of composite fermions and bosons and 
their vector-like coupling vertexes to the SM gauge 
bosons. 

The propagators of these composite particles have 
poles and residues that respectively 
represent their masses and form-factors.  
As long as their form-factors are finite, these composite particles 
behave as elementary particles. 
As discussed in Sections V and VI of Ref.~\cite{xueprd2000}, when the energy scale $\mu$
decreases to the energy threshold $\E_{\rm thre}$ and $G(\mu)\rightarrow 
G_{\rm crit}(\E_{\rm thre})$, the phase transition occurs from the 
symmetric phase to the symmetry breaking phase, 
all three-fermion and two-fermion bound states (poles) dissolve into 
their constituents, which are represented by three-fermion and two-fermion 
cuts in the energy-momentum plane, as their form-factors and binding energy 
vanish \cite{weinberg}. The propagators of these composite particles give their mass-shell conditions
\begin{eqnarray}
E_{\rm com}=\sqrt{p^2+{\mathcal M}^2}\approx {\mathcal M},\quad {\rm for }~~p\ll {\mathcal M}
\label{mcp}
\end{eqnarray}
where the mass ${\mathcal M}$ contains the negative binding energy 
$-{\mathcal B}[G(\mu)]$
and positive kinetic energies ${\mathcal K}$ of their constituents. The energy threshold 
$\E_{\rm thre}$ is determined by 
${\mathcal B}[G(\mu)]_{\mu\rightarrow\E_{\rm thre}}\rightarrow {\mathcal K}$ and vanishing form-factors of composite particles.

As required by minimizing total 
energy of the system discussed for Eq.~(\ref{bhlx}), only the 
three-fermion bound state (\ref{bound}) (top-quark channel) dissolves into a Higgs boson and a top quark (boson-fermion cut), and dynamical symmetry-breaking takes place. 
The form-factors (\ref{bound}) and (\ref{boundb}) 
$Z^{^S}_{L,R} \approx [Z^{^S}_{H}]^{1/2}[Z_{\psi}]^{1/2}$ approach to the form-factor $[\tilde Z_H]^{1/2}$ of Eqs.~(\ref{boun0}) and (\ref{thvari}), 
where $[Z_{\psi}]^{1/2}=1$ for the conventional renormalization of 
elementary fermion fields.
This means that the energy-threshold $\E_{\rm thre}$ corresponds the energy scale 
of dynamical symmetry breaking.
When the energy scale $\mu$
decreases below the energy threshold $\E_{\rm thre}$, i.e., 
$\mu < \E_{\rm thre}$, in the symmetry-breaking phase, 
the RG equations take the theory 
away from the UV fixed point towards the scaling region of the IR fixed point 
where the low-energy SM of particle physics is realized.  
On the basis of these discussions, we advocate the 
following relation for 
(i) the energy scale $\E\approx 5\,$ TeV of Eq.~(\ref{thvari}) 
extrapolated by the RG equations 
from the scaling region of 
the IR fixed point, (ii) the 
energy threshold $\E_{\rm thre}$ corresponding to the phase 
transition for dynamical symmetry breaking and (iii) the characteristic 
energy scale $\E_\xi$ setting in the scaling region of the UV fixed point
\begin{eqnarray}
\E\approx \E_{\rm thre} \lesssim \E_\xi\ll \Lambda, \quad \E\approx 5\, {\rm TeV}.
\label{scales}
\end{eqnarray}
Since $\E$ is determined by $\tilde \lambda\rightarrow 0^+$, this strongly 
indicates the occurrence of the phase transition at 
$\E\approx \E_{\rm thre}$ discussed below Eq.~(\ref{boundb}), otherwise the 
theory would run into an instability ($\tilde \lambda\sim 0^-$) beyond $\E$. 
The approximate $\E$-value (\ref{thvari}) is obtained by using the 
RG-equations (\ref{reg1}) and (\ref{reg2}), which do not give
the positively decreasing curve ``II'' of the $\beta(G)$-function 
sketched in Fig.~\ref{fixp}. 
Nevertheless, we gain some physical insight into 
the symmetry-breaking scale $\E_{\rm thre}$ and composite particle 
masses ${\mathcal M}\approx \E_\xi\gtrsim 5\, {\rm TeV}$.

Compared with the SM in the IR-stable scaling region, the composite 
field theory in 
the UV-stable scaling region has the same chiral gauge symmetries (quantum 
numbers) and couplings to gauge bosons ($\gamma,W^\pm, Z^0$ and gluon), but 
the different vector-like spectra and 1PI vertexes, 
apart from massive particles being comprised by SM elementary ones. 
These composite particles on mass-shells behave as if they were elementary, 
as long as their form-factors are finite. 
The weak and strong interactions (\ref{bhlx}) bring us into two 
distinct domains. This is reminiscent of the QCD dynamics: asymptotic free quark states near to a UV fixed point and bound hadron states near to a possible IR fixed point.

\noindent
{\bf Experiments.}
\hskip0.1cm   
These composite particles
should be produced by high-energy quarks and gauge bosons, 
if the center-of-mass energy ($\sqrt{s}\,$) of $pp$ collisions in 
the LHC is larger than their mass ${\mathcal M}$ or the threshold energy 
$\E_{\rm thre}$. These could be 
experimentally verified by possibly observing the resonances 
in the invariant masses (${\mathcal M}_{\rm inv}$) and 
kinematic distributions of 
final channels measured. We first discuss the most probable channel 
of producing the composite particles (\ref{bound}) 
of the first quark family by $pp$ collisions in the LHC. 
The elementary quarks $(u,d)_{L,R}$ are approximately massless with 
definite L- and R-chirality at TeV scales. Instead, formed Dirac 
composite particles, 
e.g., $[\bar u^{}_{Lb},(\bar u^{a}_Lu^{}_{Ra})u^b_{R}]$ or 
$[\bar d^{}_{Lb},(\bar u^{a}_Rd^{}_{La})u^b_{R}]$, are very massive, 
non-relativistic (almost static) in the center-of-mass (CM) frame. 
The most probable channel of producing them is via 
the interaction (\ref{bhlx}) of the first quark 
family, rather than via gauge interactions.      
Thus we estimate the cross-section of composite-particle formation 
$\sigma_{\rm com}\sim 1/{\mathcal M}^2$. 
If the CM energy $\sqrt{s}\gtrsim\,{\mathcal M}$ or $\E_{\rm thre}$, composite particles are not stable and 
appear as resonances (${\mathcal M}_{\rm inv}\approx {\mathcal M}$), 
and final states are two quarkonia/mesons, each of them 
decays to two jets in opposite directions (four-jets event) and 
the jet energy is about ${\mathcal M}/4$. The decay rate (inverse lifetime) 
of static composite particles $\tau^{-1}_{\rm com}\sim {\mathcal M}$ 
in the CM-frame. The quarkonia-channels $\bar u u$ and $\bar d d$ have 
the same branching ratio, which is the 
one-half of branching ratio of the meson-channel $\bar u d$. Analogously the 
bosonic composite particles (\ref{boundb}) decay to the final state of 
quarkonium or meson that forms two jets in opposite directions 
(two-jets event) and the jet energy is about ${\mathcal M}/2$. 
The same discussions apply for 
the second and third quark families, but quark pairs are most probably 
produced by two gluons with the cross-section 
$\sigma_{\rm com}\sim \alpha^2_s/{\mathcal M}^2$. The composite
particle (\ref{bound})
comprising top quark is related to the resonant channel with final states: 
a Higgs boson of energy $\sim {\mathcal M}/2$ and a $\bar t\,t$ pair, the 
latter becomes two jets of energy $\sim {\mathcal M}/4$ each, 
and three momenta are in
the same plane with almost $120^\circ$ angular separation between them, 
rather than the four-jets event for the first and second quark families.
This implies that the strong interaction (\ref{bhlx}) would give 
rise not only to
bound states, but also to peculiar kinematics of their decays, which are very 
different from the SM gauge interactions. Thus we would expect that the 
SM background should be more or less zero.   
In currently scheduled LHC runs for next 20 years, the integrated luminosity will go from 
$10\,{\rm fb}^{-1}$ up to $10^{3}\,{\rm fb}^{-1}$ and the CM energy 
$\sqrt{s}\,$ from $7
$ TeV up to $14$ TeV, then the event number of 
composite particles can be estimated 
by $\sigma_{\rm com}\times 10^{1-3}{\rm fb}^{-1}\sim 10^{5-7}$ 
for the $(u,d)$ family, $\sim 10^{3-5}$ for the $(c,s)$ 
and $(t,b)$ families, assuming ${\mathcal M}\sim 5\,$TeV. 

To end this Letter, 
we advocate that it is deserved to theoretically study the particle spectrum and symmetry 
of the strong-coupling theory (\ref{bhlx}) at the UV fixed point 
by non-perturbative numerical simulations, meanwhile experimentally verify the resonances of 
composite particles with the peculiar kinematic distributions of their final states in LHC.  

\noindent
{\bf Acknowledgment.}  
\hskip0.1cm Author is grateful to Prof.~Hagen Kleinert 
for discussions on 
the IR- and UV-stable fixed points of quantum field 
theories, to Prof.~Zhiqing Zhang for discussions on the LHC physics.

\end{document}